\documentclass[]{aa} 
 
\usepackage{graphicx}
\usepackage{xcolor}
\usepackage{natbib}  
\usepackage{scalefnt}   
\usepackage[varg]{txfonts}
\usepackage{hyperref}

\def\Ha   {$\rm H\alpha$}

\def\Cdou {$\rm ^{12}C$}
\def\Ctre {$\rm ^{13}C$}

\def\CHtre {$\rm ^{13}CH$}
\def\CNdou {$\rm ^{12}CN$}
\def\CNtre {$\rm ^{13}CN$}
\def\Cdt {$\rm ^{12}C /^{13}C$}
\def\Ou  {\ion{O}{i}}

\def\Teff  {$T_\mathrm{eff}$}
\def\logg  {$\log g$}
\def\loggf {$\log gf$}
\def\vt    {$\rm v_{t}$}

\begin{document}

\title{
\Cdt~ ratio and CNO abundances in the classical very old metal-poor dwarf HD\,140283.
\thanks{Based on observations collected  at the CFHT under program 11AB01 (PI B. Barbuy). We have also used a new ESPRESSO spectrum
and HARPS and UVES spectra from observations collected at the European Organisation for Astronomical Research in the Southern Hemisphere: Programme 0101.A-0229(A), PI M.Spite, and archives of programmes 0103.D-0118(A), PI V.Adibekian, and 080.D-0347(A), PI U.Heiter).}
}
\author{
M. Spite\inst{1}\and 
F. Spite\inst{1}\and
B. Barbuy\inst{2}
 }
\institute {
GEPI, Observatoire de Paris, PSL Research University, CNRS,
Place Jules Janssen, 92190 Meudon, France
\and
Universidade de S\~ao Paulo, IAG, Rua do Mat\~ao 1226, 
Cidade Universit\'aria, 05508-900, S\~ao Paulo, Brazil
}
\authorrunning{XXX et al.}
\titlerunning{\Cdt \  ratio in the metal-poor dwarf HD\,140283}

\abstract
{The isotope abundances provide powerful diagnostics of the chemical enrichment in our Galaxy. The star HD 140283 is one of the best-studied very metal-poor dwarf stars. It is very old, and the chemical abundance in this star is a good witness of the chemical composition of the matter in the early Galaxy. }
{The aim of this work is to measure the precise abundances of carbon, nitrogen, oxygen,
and mainly the  \Cdt ~isotopic ratio in this very old metal-poor star in order to have a good reference for the computations of the chemical evolution of the Galaxy.}
{We used very high spectral resolution data, with extremely high signal-to-noise ratios obtained
  with the spectrographs ESPaDOnS at the CFHT, ESPRESSO at the VLT, and HARPS at the ESO 3.6m telescope. 
}
{For the first time, we were able to measure the \Cdt ~ratio in a very old metal-poor dwarf that was born at the very beginning of the Galaxy: \hbox{$\rm ^{12}C /^{13}C = 33^{+12}_{-6}$}. We also obtained a precise determination of the abundance of the CNO elements in this star.
These abundances give information about the early composition of the cloud from which HD~140283 was formed. They suggest that the effect of super-asymptotic giant branch stars or fast-rotating massive stars was significant  in the early Galaxy.
}
{}
\keywords{ Stars: Abundances -- Galaxy: abundances --  Galaxy: halo}
\maketitle
%
\section{Introduction}

During the life of a galaxy,  different elements are produced inside stars of different masses and different lifetimes. The element abundances and the isotopic ratios depend on the mass of the star. Therefore elemental and isotopic abundance ratios can provide information on the age of a system and have often been used as a cosmic clock. 
The abundance of the CNO
 isotopes in metal-poor stars is indeed an important parameter for building models of galactic evolution \citep[e.g.][]{RomanoMatteucci03,ChiappiniEM08,KobayashiKU11}. \citet{RomanoMZ17,RomanoMZ19} studied the evolution of the CNO isotopic ratios in different systems with different star formation histories and different stellar initial mass functions (IMFs). For our Galaxy, these models have to explain the \Cdt~ratio in the primitive Galactic stars, in the Sun, and in the present-day interstellar medium (ISM).\\ 
The \Cdt~ratio in the Sun has been precisely measured  by \citet{AyresLL13}. From a 3D computation of the profile of the infrared (2-6 $\mu$m) rovibrational bands of carbon monoxide (CO), they measured \Cdt\, = $91.4\pm1.3$\,dex.\\
In the molecular clouds of the ISM, \citet{MilamSB05} found, based on the intensities of the hyperfine components of \CNdou~ and \CNtre~ at 113.5 and 108.8 GHz, that in the solar neighborhood, \Cdt\,$= 68\,\pm\,15$\,dex. 
Moreover, \citet{CasassusSW05} measured the \Cdt~ ratio in the diffuse ISM from absorption lines of the $\rm CH^{+}$ molecule toward nearby stars. They reported a significant variation of this ratio with a mean value of $78.3 \pm 1.8$ and a weighted rms dispersion of 12.7.
 
Only few measurements of the \Cdt~ratio in the very old primitive stars of our Galaxy have been made. Moreover, we are interested in the \Cdt~ratio in the cloud that formed the star, and this ratio in the atmosphere of the stars is strongly affected by the stellar evolution process \citep{ChanamePT05}. When a low-mass star becomes a subgiant and then a giant, the atmosphere and the deeper layers become mixed. Globally, Li and C abundances decrease, as does the \Cdt~ratio, but the N abundance increases. In the frame of the ESO large program ``First stars, first nucleosynthesis'' , the  carbon abundance was measured in a sample of extremely metal-poor (EMP) turnoff and giant stars, but the  \Cdt~ratio could only be estimated in giants. To date, it has not been measured in normal very metal-poor dwarfs or turnoff stars, whose atmospheric chemical composition is the same as the chemical composition of the cloud that formed the star. The \CHtre~ lines are indeed too weak in these stars.
  
 The \Cdt~ratio has been measured in carbon-enhanced metal-poor (CEMP) turnoff stars \citep[see e.g.][]{SivaraniBB06,BeharaBL10}. The high carbon abundance in these stars can be due to an accretion of the ejecta of a more massive companion in its AGB phase, and thus the  \Cdt~ratio does not reflect the original abundance in the cloud that formed this star.
 
The \Cdt~ratio could be estimated in the sample of EMP giants on the the lower red giant branch (RGB)  and was found to be close to 30 \citep{SpiteCH06}. These stars were called "unmixed giants" by \citet{SpiteCP05}, in comparison with the stars at the top of the RGB that have undergone a deep mixing and were called ``mixed giants''. In these  mixed giants, compared with the abundance in dwarf stars, the Li abundance has decreased by more than 100 and the C abundance by more than 10. In contrast, in the  unmixed giants on the lower RGB branch, the carbon abundance appears to have decreased by less than a factor of 2 \citep{BonifacioSC09} and the Li abundance by a factor of only about 10. It is clear, however, that these unmixed giants have undergone some mixing that might  have affected the  \Cdt~ratio in particular.
  
Because the  \Cdt~ratio in the unmixed giants is used as an important parameter when models of the chemical evolution of the Galaxy are constructed \citep{ChiappiniHM06,ChiappiniEM08,RomanoMZ17}, it is important to measure this ratio in metal-poor dwarfs or turnoff stars. In these stars, the original CNO abundances are more certain to be unaltered by mixing processes.

\begin{figure}
\label{isochrone}
\caption[]{Position of HD\,140283 (black dot) in a G vs BP-RP diagram compared to PARSEC isochrones \citep{BressanMG12}, computed for 10, 12, and 14 Gyr.}
\begin{center}
{\includegraphics [scale=0.6]{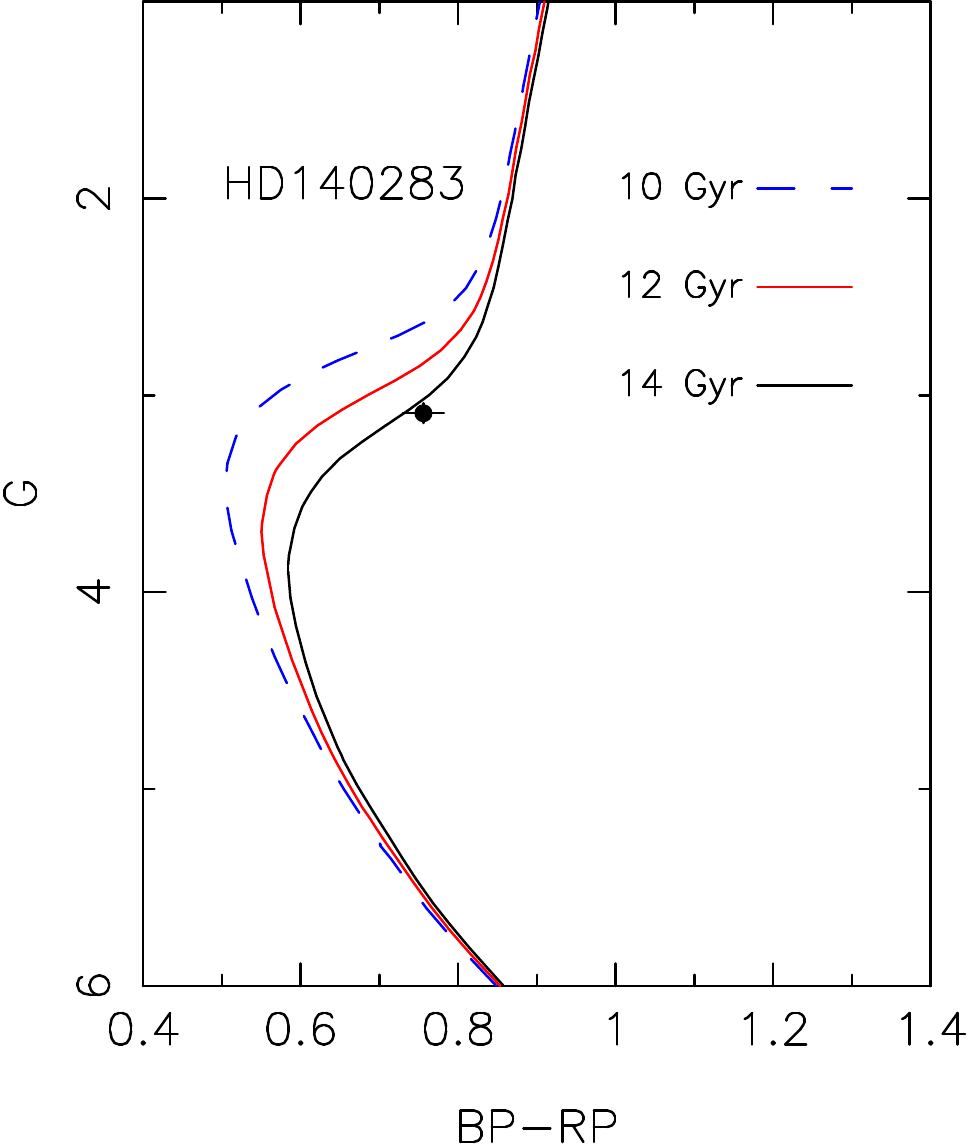}}
\end{center}
\end{figure}

\section{Nucleosynthesis of the \Ctre~ isotope.}
In the early Galaxy, the main contributors to the matter enrichment were  massive stars. They have a very short lifetime. These stars are assumed to eject a material with a high \Cdt~ratio \citep[see e.g. ][]{NomotoKT13}.  To explain the relatively low observed \Cdt~ratio, in particular, in the very metal-poor unmixed giants  \citep{SpiteCH06}, \citet{KobayashiKU11} invoked the effect of massive asymptotic giant branch (AGB) stars (with $4 M_{\odot}<M<7 M_{\odot}$) and fast-rotating massive stars \citep[FRMS)][]{MeynetEM06,ChiappiniEM08,FrischknechtHP16}.
Our paper is an attempt to measure this ratio in a classical very metal-poor star that has just passed the turnoff: HD\,140283.

\section{Main characteristics of HD\,140283}
In recent years, the old star HD 140283 has been the subject of numerous studies. The main atmospheric parameters have been determined from different characteristics of the star: excitation and ionisation equilibrium of the lines in the UV, the visible and the infrared, profiles of \Ha~ wings, and colours. All these values agree very well. All the recent determinations are gathered in Table \ref{param}, together with our adopted parameters, which are the same as were adopted by \citet{GallagherRG10} and \citet{SiqueiraBS12,SiqueiraAB15}.

HD\,140283 is a classical very metal-poor star. Its chemical composition has been studied in detail in particular by \citet{SiqueiraAB15}. They found a metallicity [Fe/H]=--2.59.\footnote{For each element X, we adopted the classical notations~~ A(X)=(log (N(X)/N(H)) + 12),~~~~$\rm [X/H]=A(X)_{\star}-A(X)_{\odot}$ ~~, and~~~ [X/Fe]=[X/H]--[Fe/H].}  

It is a very old star that formed shortly after the Big Bang. From isochrones computed using the 2012 version of the University of Victoria code \citep{VandenBergBD12} with a parallax obtained from the Fine Guidance Sensors of the Hubble Space Telescope ($\rm\pi = 17.15 \pm 0.14\,mas$), \citet{BondNVdB13} found an age of $14.46 \pm 0.31$ Gyr.\\ 
This parallax is larger than the more precise measurement of the Gaia EDR3  
\citep[][]{GaiaEDR3-Brown}: $\rm\pi = 16.30 \pm 0.03 \,mas$  (after a 0.03 offset correction \citep{LindegrenHB18}). 
In \hbox{Fig. 1} we present the position of HD\,140283 in a G versus BP-RP diagram. The G magnitude and the  BP-RP colour given by the Gaia-DR2 were corrected for the distance and the reddening of the star using the map STILISM\footnote{\tt https://stilism.obspm.fr} \citep{LallementCRD18}. The error in G is negligible and the error on BP-RP is mainly due to the uncertainty in the reddening. The position of HD\,140283 in Fig. 1 is compared to the PARSEC isochrones \citep{BressanMG12} computed for 10, 12, and 14 Gyr. 
This diagram confirms that the age of HD\,140283 is very close to 14 Gyr.\\

\begin{table}
\begin{center}   
\caption[]{Recent stellar parameters for HD 140283 in the literature 
and adopted values}
\label{param}
\begin{tabular}{rccclcrrrrrrr}
\hline
\noalign{\vskip 0.05cm}
\Teff &\logg & [Fe/H] &\hbox{\vt } &Reference \\
\noalign{\vskip 0.05cm}
\noalign{\hrule}
 5750 &  3.40 & --2.50 & 1.4 &\citet{AokiIK04}\\
 5753 &  3.70 & --2.40 & 1.5 &\citet{AsplundLN06}\\
 5769 &  3.73 & --2.54 & 1.5 &\citet{HosfordRG09}\\ 
 5812 &  3.75 &   -   &  -  &\citet{CasagrandeRM10}\\
 5600 &  3.66 & --2.67 & 1.2 &\citet{Roederer12}\\ 
 5700 &  3.60 & --2.60 & 1.3 &\citet{PetersonKur15} \\  
 5650 &  3.40 & --2.70 & 1.7 &\citet{AfsarSF16}\\
\hline
\multicolumn{3}{l}{Adopted parameters}\\
\Teff &\logg & [Fe/H] &\hbox{\vt }\\
 5750 & 3.70 & --2.59  & 1.4 &\citet{SiqueiraAB15}\\ 
 \hline
\end{tabular} 
\end{center}  
\end{table}

\section{Observations} \label{sec:obs}
HD\,140283 was observed in 2011 within the program 11AB01 (PI: B. Barbuy) at the Canada-France-Hawaii Telescope (CFHT) with the spectrograph ESPaDOnS in order to obtain a high-resolution spectrum with a very high signal-to-noise ratio (S/N) in the wavelength range 3700\AA~- 10475\AA. The spectrum was reduced using the CFHT pipeline {\tt Upena} (more information about the reduction procedure is given by \citet{SiqueiraAB15}).
The resolving power of the resulting spectrum is $R=81000,$ and the S/N per pixel is about 1500 at 4200\AA.

A very good spectrum of HD\,140283 was recently also obtained with the ESPRESSO spectrograph, which is installed at the incoherent combined Coud\'e facility of the Very Large Telescope. The resolving power of this spectrum,  140\,000 at 4200\AA, is  higher than the resolving power of the CFHT spectrum, but the S/N is lower: about 700. \\
The full width at half maximum (FWHM) of the lines in this spectrum is about 0.06 {\rm \AA} and the size of a pixel is 0.005\,{\rm \AA}. Following \citet{Cayrelformule88}, it is thus possible to measure lines as weak as 0.04\,m{\rm \AA} on this spectrum. This theoretical formula is rather optimistic (it neglects, in particular, the uncertainty on the location of the continuum), therefore observers generally multiply this estimate by 2. To be on the safe side of the estimation, we conclude that it is possible to detect lines as weak as 0.1\,m{\rm \AA}. This value is also a good estimate of the error on the equivalent widths of the weak lines. The same Cayrel formula applied to the ESPaDOnS spectrum leads to about the same estimate of the error (the higher S/N of this spectrum compensates for its lower resolution).

The ESO archive also contains a HARPS spectrum with a resolving power  $R=120000,$ but the S/N is only 300 around 4200\AA. This HARPS spectrum was used to check the compatibility of our results.


\begin{table}
\begin{center}
\scalefont{0.85}
\caption[]{Abundance ratios of the C, N, O elements in HD\,140283, with [Fe/H]=--2.57} 
\label{tab:abCNO}
\begin{tabular}{c@{~~~}c@{~~~}c@{~~~}c@{~~~}c@{~~~}c@{~~~}c@{~~~}c@{~~~}c@{~}r@{~~~~~~~~}c@{~}r@{~~~~~~~~}c@{~}c@{~}r@{~~~~~~~}c@{~}r@{~~~~~~~~}c@{~}c@{~}r@{~~~~~~~}c@{~}r@{~}}
\hline
[C/Fe] &   [12C/13C]& [12C/13C]      &  [N/Fe]*  & [O/Fe]** &  [C/O] & [N/O]\\
       &     number &  mass fraction &          &   NLTE &               \\
+0.41  &        33  &       31       &  +0.16   &  +0.79 &  -0.39 & -0.63\\         
\noalign{\vskip 0.2cm}
\multicolumn{7}{l}{*The N abundance is corrected by --0.4\,dex }\\ 
\multicolumn{7}{l}{~~~(shift of the Kurucz log gf values of the NH band),}\\
\multicolumn{7}{l}{**The oxygen abundance is corrected by --0.1\,dex for NLTE.}\\
\hline
\end{tabular} 
\end{center}  
\end{table}

\begin{figure*}
\label{fig:spec}
\begin{center}   
\caption[]{Observed ESPaDOnS (CFHT) ESPRESSO  and HARPS (ESO) spectra (crosses) in the region of the \Ctre~ features.  Synthetic spectra (thin blue lines) are shown for A(\Cdou)=6.31, and A(\Ctre)=--5 (no \Ctre), A(\Ctre)=4.61 ( $\rm ^{12}C /^{13}C = 50$), and A(\Ctre)=5.01 ( $\rm ^{12}C /^{13}C = 20$).  The red line represents the best fit.}
{\includegraphics [scale=0.70]{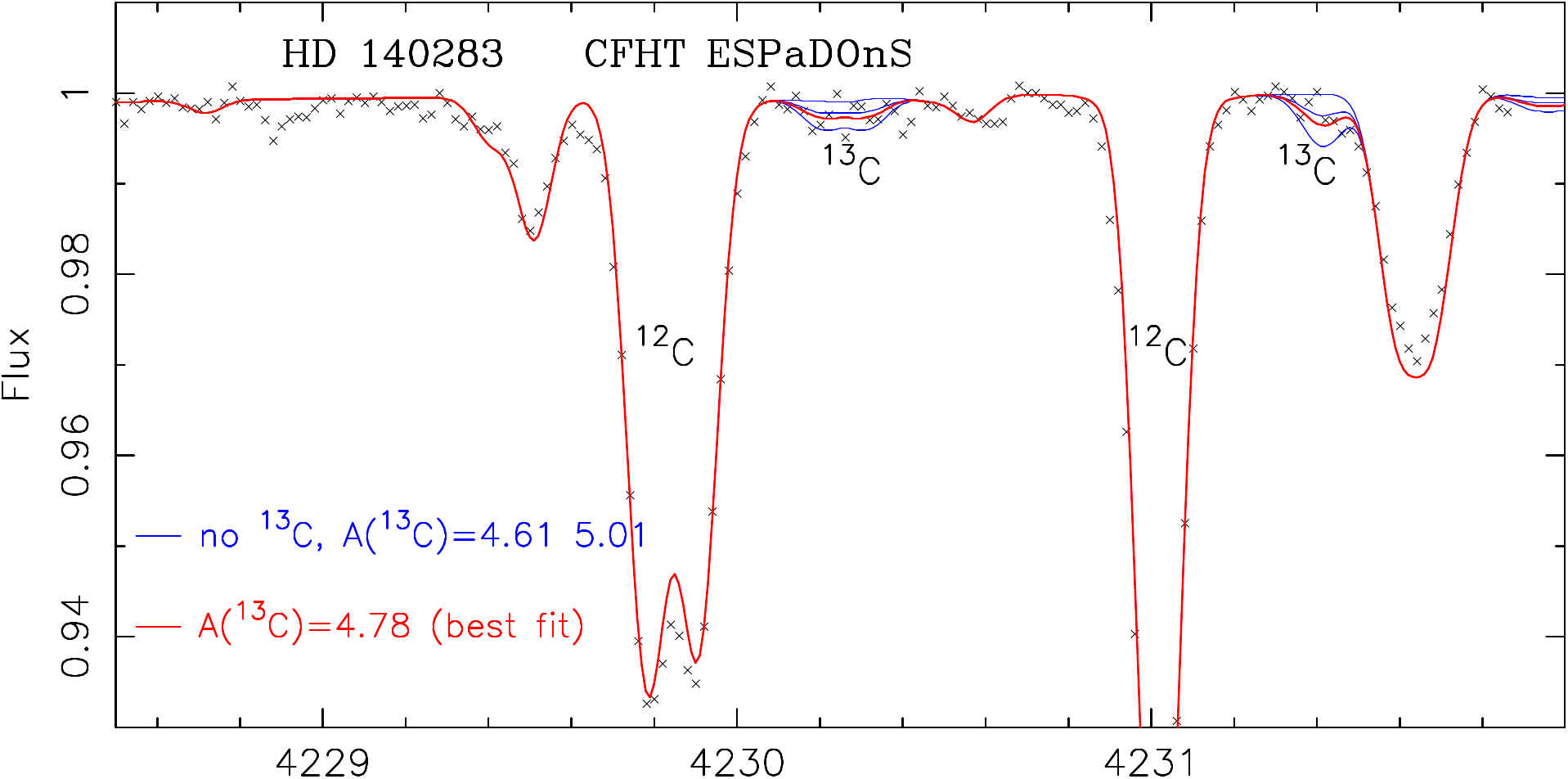}}
{\includegraphics [scale=0.70]{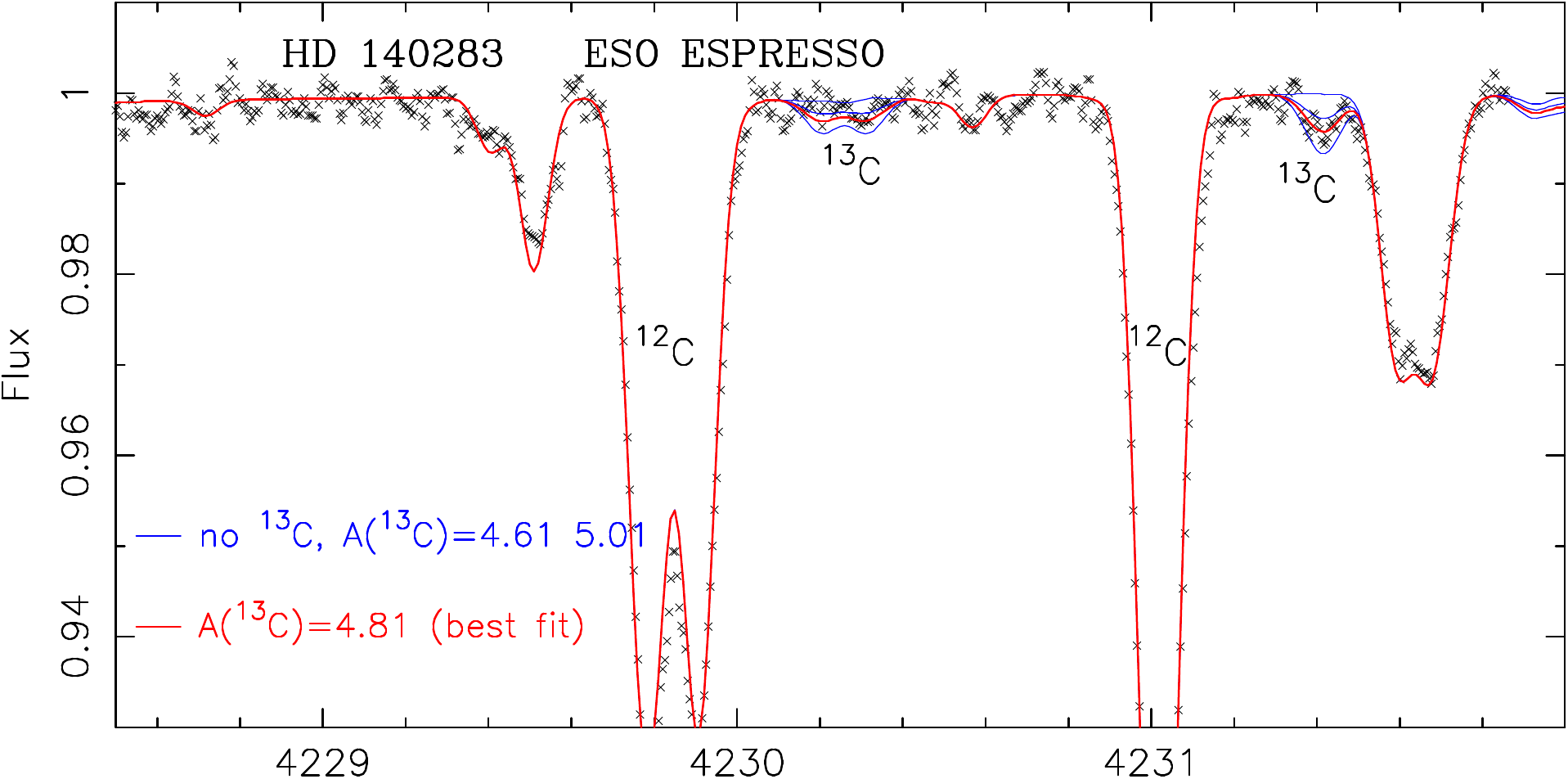}}
{\includegraphics [scale=0.70]{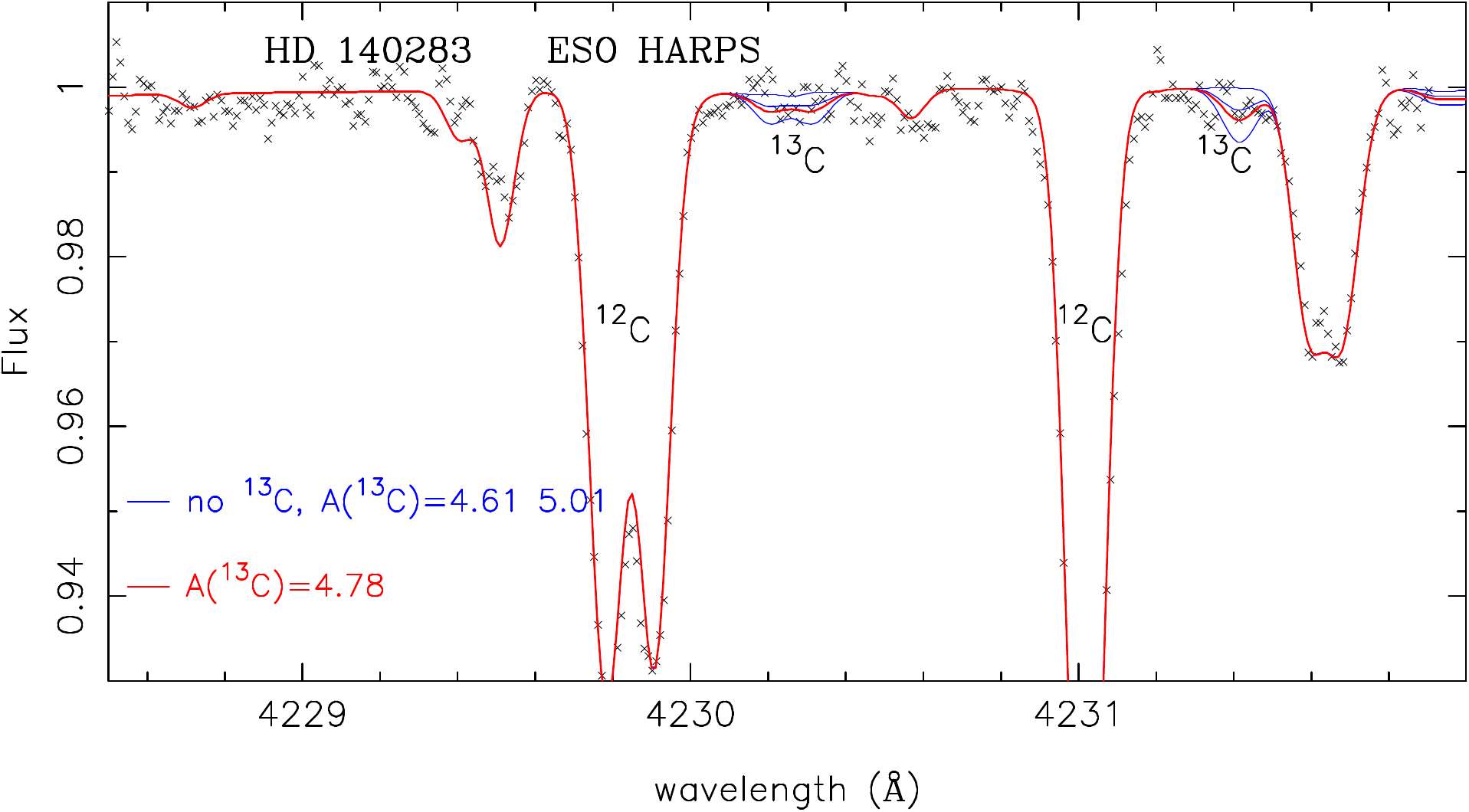}}
\end{center}
\end{figure*}

\section{Determination of C, \Cdou,~ and \Ctre~ abundances}

We  interpolated an atmosphere model in the MARCS grid \citep{GustafssonBE75,GustafssonEE03,Gustafsson08,Plez08} with the parameters defined in Table \ref{param}, and in Fig. 2 we show a synthetic spectrum in the region of the G carbon-band  with the current version of the {\tt turbospectrum} code \citep{AlvarezP98}.\\
We made use of the updated line list of the \Cdou~ and \Ctre~ bands given by \citet{MasseronPVE14}. A major improvement in these data was achieved because they are the result of a global fit of astronomical observations and  laboratory measurements for the three electronic transitions A--X, B--X, and C--X. 
Several features of the \Ctre~band exist, but the stronger \CHtre~lines are at 4230.3 and 4231.45\,\AA.
The best fit was obtained for A(\Cdou)\,=\,6.31  and A(\Ctre)\,=\,4.80 (mean of the ESPaDOnS and ESPRESSO values,
Fig. 2).\\
These values correspond to  A(C)\,=\,6.32, or, with $\rm A(C)_{\odot}=8.50$ \citep{CaffauLS09},  [C/H]\,=\,--2.18, and thus [C/Fe]\,=\,+0.41. This agrees well with the value found by \citet{SiqueiraAB15}, [C/Fe]\,=\,+0.46. This value of [C/Fe] is also very close to the mean value found in the turnoff very metal-poor stars,  $\rm[C/Fe] = +0.45  \pm 0.10$ \citep{BonifacioSC09}.\\
The  \Ctre ~features are well fitted with A(\Ctre)\,=\,4.80, 
  and the equivalent width of the \Ctre~ feature at 4231.4\,m{\rm \AA}
  (measured on the synthetic spectrum) 
is close to 0.4\,m{\rm \AA}. If we adopt an error of 0.1\,m{\rm \AA} on the equivalent width (see section \ref{sec:obs}), this error therefore corresponds to 
\hbox{$\rm A(^{13}C) = 4.80^{+0.09}_{-0.13}$}  or 
\hbox{$\rm ^{12}C /^{13}C = 33^{+12}_{-6}$}.

\section{Nitrogen and oxygen abundance}

\begin{figure*}
\label{fig:NHOxy}
\begin{center}   
\caption[]{Top panel: NH band in the UVES spectrum computed with A(N)=5.7 and A(N)=6.3 (blue lines) and with A(N)=5.85 (best fit, red line). Bottom panel: \Ou~ triplet in the ESPRESSO spectrum. The synthetic spectra have been computed with A(O)=6.7, 7.3 (blue lines) and the best fit A(O)=7.06 (red line). The crosses represent the observed spectra. The abundances are raw values.  The N abundances must be corrected by --0.4 (shift of the Kurucz log gf values) and the oxygen abundance by --0.1 (NLTE).
}
{\includegraphics [scale=0.70]{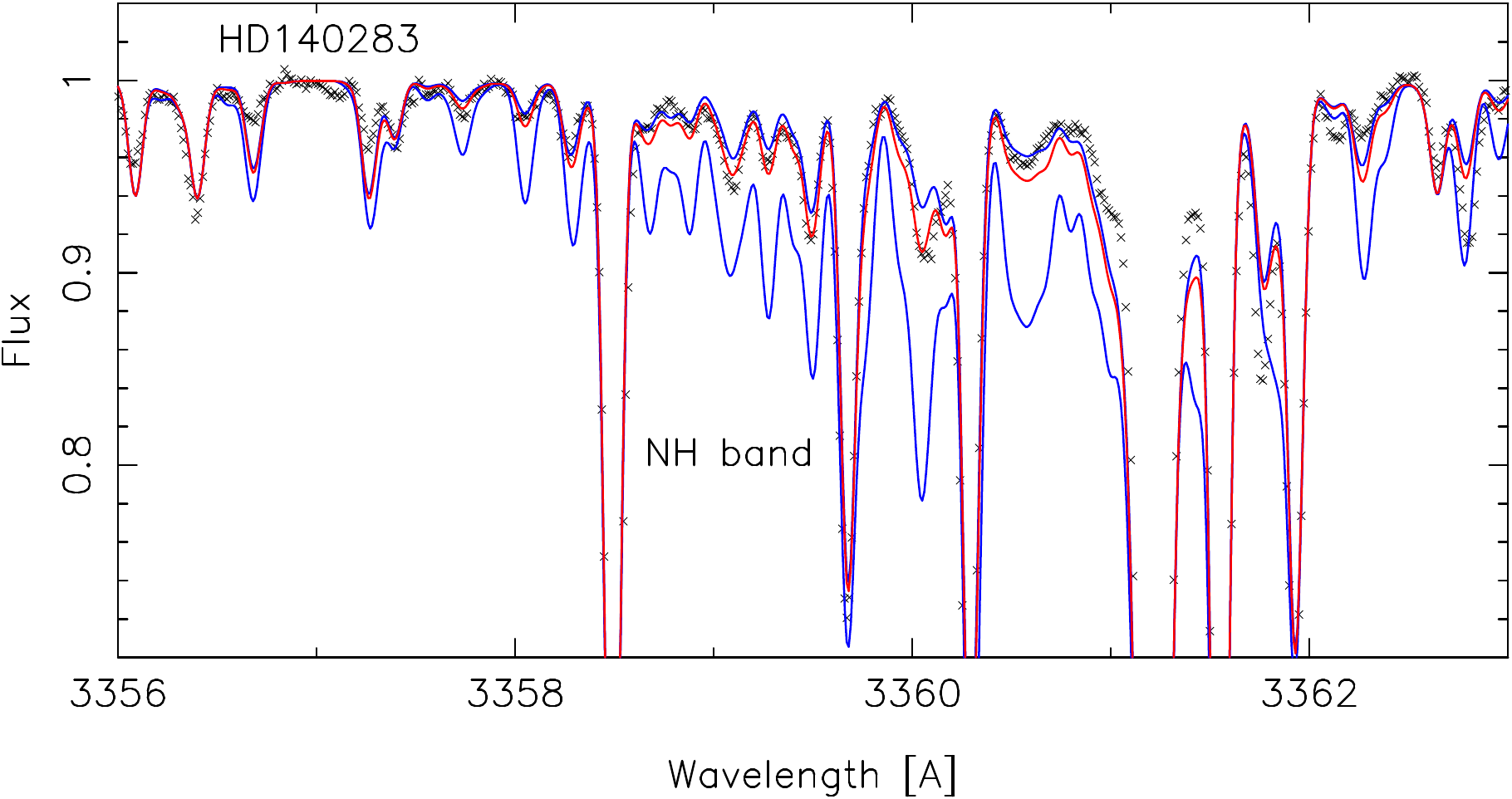}}
{\includegraphics [scale=0.70]{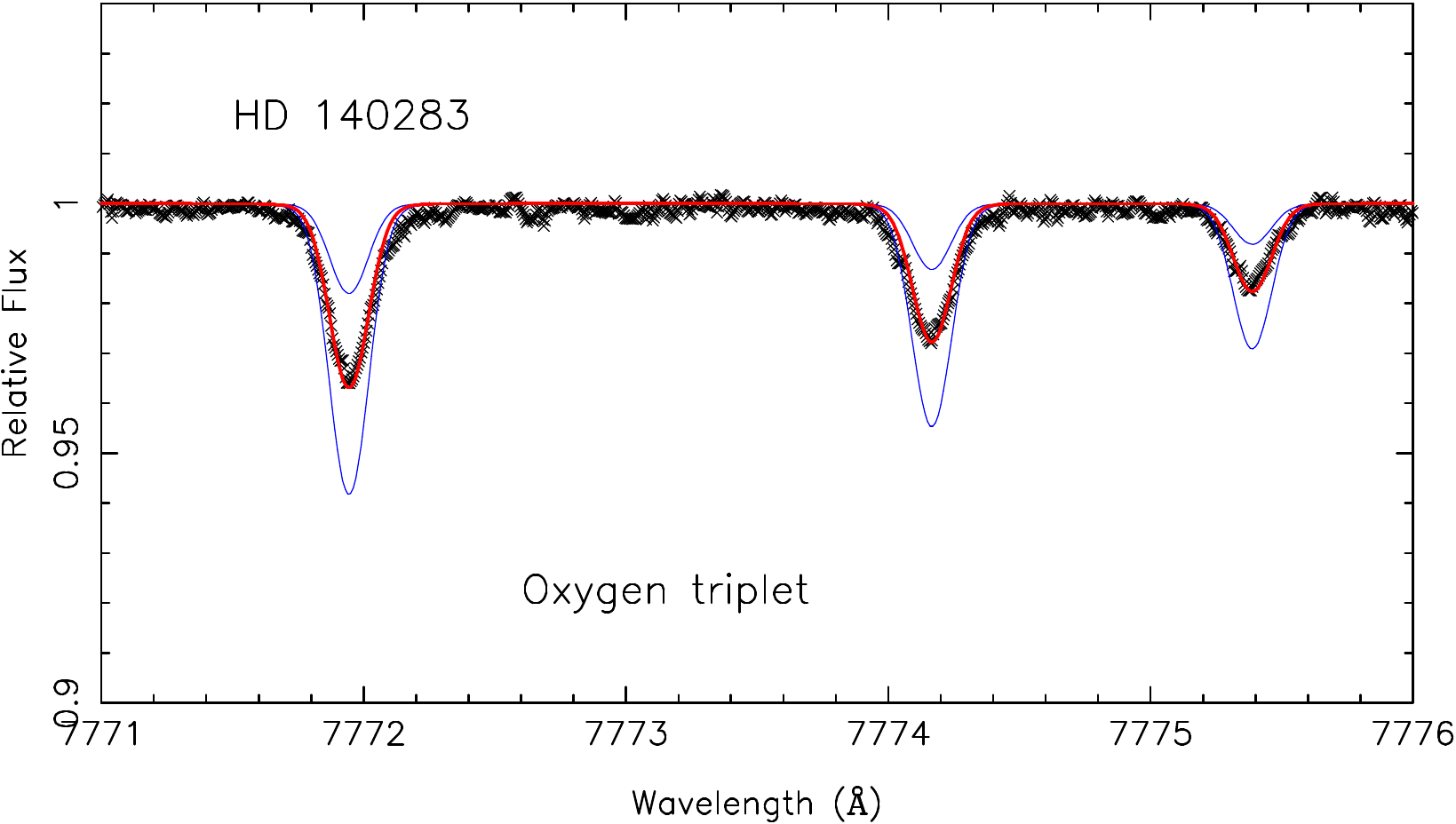}}
\end{center}
\end{figure*}

From the profile of the CN band at 3883\AA~, \citet{SiqueiraAB15} derived a very high value of the nitrogen abundance A(N)\,=\,6.30 and thus [N/Fe]\,=\,+1.06. Because in dwarf stars, the CN band is very weak, a very small change in the position of the continuum induces a large variation of the nitrogen abundance. We thus decided to determine the nitrogen abundance  from the NH band in the near UV from a recently obtained UVES spectrum 
(Fig. 3, top panel).
The line list and the molecular data of the NH band comes from Kurucz. The best fit was obtained with A(N)\,=\,5.85 or [N/H]\,=\,--2.01 and [N/Fe]\,=\,+0.56 (adopting $\rm A(N)_{\odot}=7.86$ following \citet{CaffauLS09}).\\
  \citet{SpiteCP05} showed that there is a systematic difference between the N abundance derived from the NH or the CN bands based on the Kurucz data:  $\rm A(N)_{NH} - A(N)_{CN} =0.4$ dex. It is generally admitted that the \loggf~ values in the NH data of Kurucz are 0.4 dex too low. This correction was also applied by \citet{PasquiniEB08} in globular clusters stars. In table \ref{tab:abCNO} we list the corrected value.\\
With this CN scale, the N abundance in HD\,140283 is even lower: A(N)\,=\,5.45 and [N/Fe]\,=\,+0.16. \\
In Fig. 3, the value 
A(N)\,=\,6.3 \citep{SiqueiraAB15}  is incompatible with the observed spectrum of the NH band. On the other hand, we verified that the  low value of the N abundance found from the NH band is compatible with the observed CN profile.\\

The oxygen abundance was also computed from the ESPRESSO spectrum around 7770\,\AA. We found A(O)\,=\,7.06  (Fig. 3, bottom panel), 
which perfectly agrees with the value deduced from the ESPaDOnS spectrum by \citet{SiqueiraAB15}. This value was corrected by --0.1 dex for non-local thermodynamic equilibrium (NLTE) following \citet{ZhaoMashon16}. As a consequence, in this star \citep[with $\rm A(O)_\odot = 8.76$ following ][]{CaffauLS09,Steffen15},  [O/H]\,=\,--1.76 and [O/Fe] = +0.77.

\section{Discussion and conclusion}  
For the first time, we have measured the precise value of the $\rm ^{12}C /^{13}C $ ratio in a very old galactic metal-poor dwarf HD\,140283. 
We also derived precise oxygen and nitrogen abundances in this star. The abundance ratios of the C, N, O elements are summarised in Table \ref{tab:abCNO}.\\
At the evolution stage of HD\,140283, the star has not undergone any mixing with the deep layers. In particular, the Li abundance, which is more sensitive to mixing processes than the C and N elements, is normal in this star according to \citet{SiqueiraAB15}:~A(Li) = 2.14,  a value very close to the ``plateau''. (In the low RGB stars, which have until now been used as a reference for the abundance of the CNO elements in the early Galaxy, the Li abundance is more than ten times lower than in the dwarfs).\\
As a consequence, in  HD\,140283 the C, N, O abundances and the isotopic $\rm ^{12}C /^{13}C $ ratio should be the same as the original values in the cloud that formed the star. The [C/Fe] ratio in HD\,140283 (+0.41) perfectly agrees with the mean value found by \citet{BonifacioSC09} in a sample of very and extremely metal-poor dwarfs and turnoff stars. The [O/Fe] ratio in HD\,140283 also agrees well with the value measured by \citet{GonzalezHernandezBL08} in the EMP dwarf CS\,22876-032. 
Moreover, the ratios [C/Fe], [N/Fe], [O/Fe], [N/O], [C/O], and   $\rm ^{12}C /^{13}C $ in HD\,140283 are not very different from the mean value found  in the low  RGB stars called unmixed giants in \citet{SpiteCP05} (see their Figs. 6, 11, and 14). The first dredge-up does not seem to have severely affected the CNO abundance ratios and in particular the ratio $\rm ^{12}C /^{13}C $.

HD\,140283 is a very old star whose atmosphere is assumed to be a good witness of the original abundances of the CNO elements. We measured  $\rm ^{12}C /^{13}C = 32$ (or 30 in mass fraction), a rather low ratio. Because the classical massive supernovae are not able to eject a significant amount of \Ctre~ \citep[see e.g. ][]{NomotoKT13}, this low $\rm ^{12}C /^{13}C$ ratio suggests that the material that forms the very old galactic stars has been enriched by FRMS
or massive AGB stars.\\
In FRMS, rotation triggers the production of primary \Ctre. It indeed allows the diffusion of \Cdou~that is produced in the He-burning zone into the H-burning region, where it is converted into \Ctre~ by the CNO cycle.\\ 
Another possibility is an enrichment of the matter by super-AGB or massive AGB stars. This could be an additional source of \Ctre~in the early stages of the Galaxy formation \citep{DohertyGL14,Gil-PonsDG21}.


0
\begin {acknowledgements} 
BB acknowledges partial financial support from Fapesp, CNPq, and CAPES.
This work uses results from the European Space Agency (ESA) space
mission Gaia.  Gaia data are being processed by the Gaia Data
Processing and Analysis Consortium (DPAC).  Funding for the DPAC is
provided by national institutions, in particular the institutions
participating in the Gaia MultiLateral Agreement (MLA).  The Gaia
mission website is https://www.cosmos.esa.int/gaia.  The Gaia archive
website is https://archives.esac.esa.int/gaia.  
\end{acknowledgements}

\bibliographystyle{aa}
{}

\end{document}